\documentclass[%
reprint,
superscriptaddress,
 amsmath,amssymb,
 aps,
prb,
]{revtex4-1}

\usepackage {graphicx,epsfig,graphics,color}% Include figure files
\usepackage{dcolumn}% Align table columns on decimal point
\usepackage{bm}% bold math
\usepackage{hyperref}% add hypertext capabilities
\usepackage{sidecap}
\usepackage{float}

\begin{document}
\title{Flat-band spin dynamics and phonon anomalies of the saw-tooth spin-chain system Fe$_2$O(SeO$_3$)$_2$}

\author{V. P. Gnezdilov}
\affiliation{B.I. Verkin Institute for Low Temperature Physics and Engineering, NASU, 61103 Kharkiv, Ukraine}
\affiliation{Institute for Condensed Matter Physics, TU Braunschweig, D-38106 Braunschweig, Germany}

\author{Yu. G. Pashkevich}
\affiliation{O.O. Galkin Donetsk Institute for Physics and Engineering, NASU, Kyiv-Kharkiv, Ukraine}

\author{V. S. Kurnosov}
\affiliation{B.I. Verkin Institute for Low Temperature Physics and Engineering, NASU, 61103 Kharkiv, Ukraine}

\author{O. V. Zhuravlev}
\affiliation{O.O. Galkin Donetsk Institute for Physics and Engineering, NASU, Kyiv-Kharkiv, Ukraine}

\author{D. Wulferding}
\affiliation{Institute for Condensed Matter Physics, TU Braunschweig, D-38106 Braunschweig, Germany}
\affiliation{Laboratory for Emerging Nanometrology (LENA), TU Braunschweig, D-38106 Braunschweig, Germany}

\author{P. Lemmens}
\affiliation{Institute for Condensed Matter Physics, TU Braunschweig, D-38106 Braunschweig, Germany}
\affiliation{Laboratory for Emerging Nanometrology (LENA), TU Braunschweig, D-38106 Braunschweig, Germany}

\author{D. Menzel}
\affiliation{Institute for Condensed Matter Physics, TU Braunschweig, D-38106 Braunschweig, Germany}

\author{E. S. Kozlyakova}
\affiliation{M.V. Lomonosov Moscow State University, 119991 Moscow, Russia}

\author{A. Yu. Akhrorov}
\affiliation{M.V. Lomonosov Moscow State University, 119991 Moscow, Russia}

\author{E. S. Kuznetsova}
\affiliation{M.V. Lomonosov Moscow State University, 119991 Moscow, Russia}

\author{P. S. Berdonosov}
\affiliation{M.V. Lomonosov Moscow State University, 119991 Moscow, Russia}

\author{V. A. Dolgikh}
\affiliation{M.V. Lomonosov Moscow State University, 119991 Moscow, Russia}

\author{O. S. Volkova}
\affiliation{M.V. Lomonosov Moscow State University, 119991 Moscow, Russia}
\affiliation{National University of Science and Technology MISiS, 119049 Moscow, Russia}
\affiliation{National Research South Ural State University, 454080 Chelyabinsk, Russia}

\author{A. N. Vasiliev}
\affiliation{M.V. Lomonosov Moscow State University, 119991 Moscow, Russia}
\affiliation{National University of Science and Technology MISiS, 119049 Moscow, Russia}
\affiliation{National Research South Ural State University, 454080 Chelyabinsk, Russia}

\date{\today}

\begin{abstract}

Fe$^{3+}$ $S = 5/2$ ions form saw-tooth like chains along the $a$ axis of the oxo-selenite Fe$_2$O(SeO$_3$)$_2$ and an onset of long-range magnetic order is observed for temperatures below $T_C = 105$ K. This order leads to distinct fingerprints in phonon mode linewidths and energies as resolved by Raman scattering. In addition, new excitations with small linewidths emerge below $T = 150$ K, and are assigned to two-magnon scattering processes with the participation of flat-band and high energy magnon branches. From this a set of exchange coupling constants is estimated. The specific ratio of the saw-tooth spine-spine and spine-vertex interactions may explain the instability of the dimer quantum ground state against an incommensurate 3D magnetic order.

%\begin{description}

%\item[PACS numbers]{XXX, YYY, ZZZ}
%\pacs{XXX, YYY, ZZZ}

%\end{description}
\end{abstract}

\maketitle

\section{Introduction}

There is great interest in magnetic quantum systems with well known magnetic ground states and complex excitation spectra. One such system is the antiferromagnetic saw-tooth chain, or $\Delta$ chain -- a one-dimensional lattice spin system with a topology of corner-sharing antiferromagnetically connected triangles. In contrast to many other frustrated systems, the ground state of a perfect $S = 1/2$ saw-tooth chain with couplings between the base ($J_{bb}$) and the base-vertex neighbor-nearest (NN) spins ($J_{bv}$) is exactly known~\cite{kubo-93, nakamura-96, sen-96}. It consists of a two-fold degenerate superposition of spin-singlets on either each left pair or each right pair of spins. Such a $J_{bb} = J_{bv} = J$ saw-tooth chain has a dispersionless small gap $\Delta = 0.215 J$~\cite{nakamura-96, sen-96, blundell-03} with the lowest excitations created by ``kink'' and ``anti-kink'' pairs. These are topological defects with spin $S = 1/2$. Further studies revealed a variety of ground states considering various ratios of $J_{bb}/J_{bv}$~\cite{blundell-03}, Ising and Heisenberg bonds and spin-lattice interaction~\cite{ohanyan-09, bellucci-10}, Dzyaloshinskii-Moriya interactions (DMI)~\cite{hao-11}, or different spin states on the corners of triangles~\cite{chandra-04}.

The saw-tooth lattice topology implies the presence of magnetic ions in at least two different Wyckoff positions. Thus, an interplay between different magnetic order parameters and geometrical frustration may yield a rich magnetic phase diagram. An easy switching between different 3D ordered ground states requires a complex and unusual spin dynamics with the presence of few low energy magnon modes. The latter modes might originate from zero energy flat-band modes activated by exchange interactions which are slightly different from characteristic interactions of the ideal saw-tooth structure. In the classical limit some saw-tooth model systems possess zero energy flat-band modes, remarkably similar to a Kagome antiferromagnet~\cite{zhitomirsky-04,zhitomirsky-05}. Strongly correlated flat-band systems with localized magnon states in high magnetic fields are the subject of intensive studies (for a review see also Ref.~\cite{derzhko-15}), as flat-band modes offer promising possibilities for manipulating the propagation of waves of any origin~\cite{bergholtz-13, leykam-18, vicencio-15}. Therefore, they are of great interest for the development of magnonic devices~\cite{wang-18}. For a device based on a saw-tooth photonic lattice a dedicated inhibition of transport has recently been reported together with a highly degenerate flat-band~\cite{weimann-16}. A high energy magnon flat-band mode has been observed in a Kagome lattice ferromagnet in the absence of magnetic fields~\cite{chisnell-15}.

Despite this theoretical effort, only a small number of respective experimental realizations exists. This includes the delafossite YCuO$_{2.5}$~\cite{cava-93, bacq-05}, olivines Zn$L$$_2$S$_4$ ($L$ = Er, Tm, Yb)~\cite{lau-06}, Mn$_2$GeO$_4$~\cite{white-12, honda-12, honda-17}, euchroite Cu$_2$(AsO$_4$)(OH)$\cdot$3H$_2$O~\cite{kikuchi-11}, CuFe$_2$Ge$_2$~\cite{may-16}, and oxy-arsenate Rb$_2$Fe$_2$O(AsO$_4$)$_2$~\cite{garlea-14}. These compounds have been found to be of saw-tooth chain type realizing different cases of complex magnetism. Experimentally, no long-range magnetic order has been found in YCuO$_{2.5}$, Zn$L$$_2$S$_4$, and Cu$_2$(AsO$_4$)(OH)$\cdot$3H$_2$O even at low temperatures. In Cu$_2$(AsO$_4$)(OH)$\cdot$3H$_2$O a large spin gap of about $\Delta_0 = 90 \pm 5$ K and a new mid-gap of about $\Delta_M = 56$ K have been found in NMR and high field magnetization data, respectively.

With respect to 3D ordered ground states the most prominent representative is Mn$_2$GeO$_4$ with $T_{N1} = 47$ K and additional first-order magnetic transitions at $T_{N2} = 17$ K and $T_{N3} = 5.5$ K~\cite{white-12, honda-12, honda-17}. Here, the double-$Q$ magnetic structure in the multiferroic state can be regarded as a canted conical spiral spin state. The saw-tooth compound Rb$_2$Fe$_2$O(AsO$_4$)$_2$ orders antiferromagnetically (AFM) below $T_N = 25$ K. Here, an external magnetic field leads to a transition from an AFM to a ferrimagnetic state. Neutron diffraction on CuFe$_2$Ge$_2$ revealed AFM order below $\approx 175$ K and an incommensurate spin structure (IC) below $\approx 125$ K~\cite{may-16}.

Here we report a Raman scattering study accompanied by magnetic susceptibility measurements of a new saw-tooth chain compound -- the iron oxoselenite Fe$_2$O(SeO$_3$)$_2$. Unlike other saw-tooth chain compounds, its magnetic susceptibility evidences low-dimensional behavior above $T_C \sim 105$ K which can not be described by a Curie-Weiss law. This highlights the proximity of Fe$_2$O(SeO$_3$)$_2$ to a quantum critical point (QCP). Below $T_C$ the susceptibility follows antiferromagnetic behavior. We observe several phonon anomalies related to the onset of 3D magnetic order in Fe$_2$O(SeO$_3$)$_2$. Our lattice dynamic calculations support magnetoelastic coupling between lattice and magnetic subsystems. Below $T_C$ we observe a few highly polarized and relatively narrow magnetic excitations at different energy scales. We assign these signals to two-magnon scattering with the participation of weakly dispersive magnon branches. We uncover four-fold and two-fold degenerate modes which are almost flat on the $k_yk_z$ plane of reciprocal space, independently on the presence of 3D interchain exchange interactions. Furthermore, we demonstrate that 3D magnetic order can develop from a fluctuating valence bond state as a consequence of specific relations between exchange interactions.

\section{Experimental details}

Raman scattering experiments were performed with different setups in quasi-backscattering geometry on the as-grown surfaces of Fe$_2$O(SeO$_3$)$_2$ single crystals. We used an excitation wavelength $\lambda = 532.1$ nm of a Nd:YAG solid-state laser with the power $P = 2$ mW. The scattered light was collected and dispersed by a triple monochromator DILOR XY spectrometer onto a liquid-nitrogen-cooled Horiba Jobin Yvon, Spectrum One CCD-3000V detector. The temperature dependence of the Raman spectra was obtained in a variable temperature closed-cycle cryostat (Oxford/Cryomech Optistat). A Horiba LabRam HR800 spectrometer was used for measurements from the top of the needle-like sample with a regular size of $0.1 \times 0.1 \times 1$ mm$^3$. Raman spectra were measured in $aa$, $bb$, $cc$, $ac$, and $bc$ scattering geometries.

\section{Experimental Results}

Fe$_2$O(SeO$_3$)$_2$ crystallizes in the orthorhombic structure~\cite{giester-96} (space group $Pccn$, \# 56, $Z = 8$) with the unit cell constants $a = 6.571(4)$ {\AA}, $b = 12.83(1)$ {\AA} and $c = 13.28(1)$ {\AA}. The structure consists of triangle-based magnetic chains made of $S = 5/2$ Fe$^{3+}$ ions occupying three crystallographically distinct sites (Fe1, Fe2, and Fe3). Every tooth of these chains is centered by one oxygen atom, coordinating two octahedra Fe1O$_6$ and Fe3O$_6$ and one tetrahedron Fe2O$_4$. The Fe3 clusters that are connected to each other via Fe1 and Fe2 corners form the chains along the $a$ axis with teeth alternatively tilted along different diagonals in the $bc$-plane as shown in Fig. 1(b). The triangular Fe clusters sharing Fe1 and Fe2 corners form the saw-tooth like chains, with their teeth alternatively tilted from the $ac$-plane.

\begin{figure*}
\label{figure1}
\centering
\includegraphics[width=16cm]{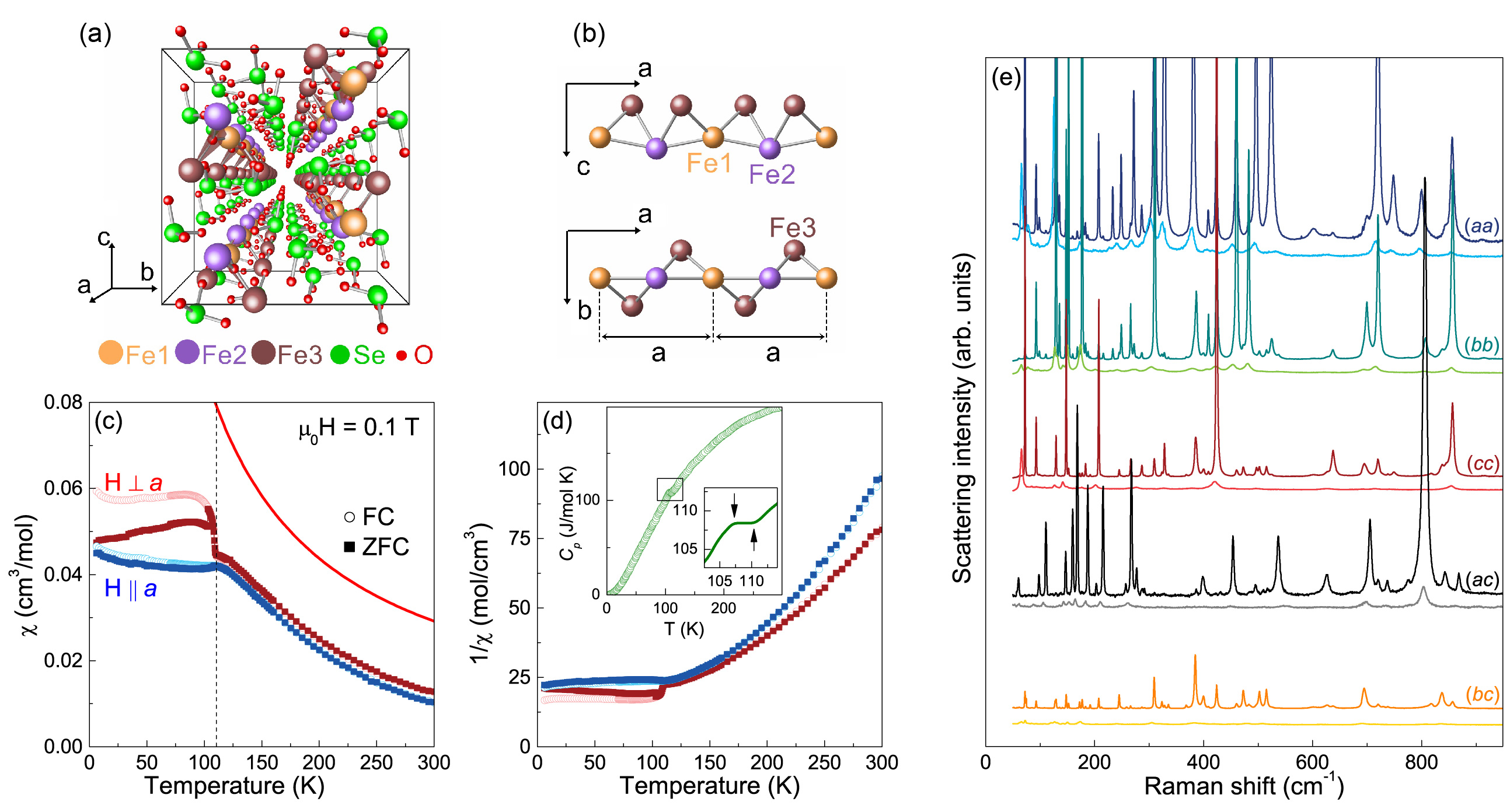}
\caption{(Color online) (a) Crystal structure of Fe$_2$O(SeO$_3$)$_2$. Four saw-tooth chains in the primitive cell running along the $a$ axis. (b) Saw-tooth arrangement of the Fe$^{3+}$ ions. Basic triangular clusters are formed by three Fe$^{3+}$ ions. Projections onto the $ab$ plane and the $ac$ plane of one representative saw-tooth chain are shown. The primitive unit cell contains 16 iron ions, which include one Fe1, one Fe2 and two Fe3 ions for each of the four chains. (c) Temperature dependence of the magnetic susceptibility $\chi$ obtained in field-cooled (open circles) and zero-field-cooled (solid squares) modes. The red line represents the Curie-Weiss law for non-interacting Fe$^{3+}$ ions ($S=5/2$, $\mu = 5.92$ $\mu_B$). (d) Inverse magnetic susceptibility. The inset shows the temperature dependence of the specific heat capacity. (e) Raman spectra of Fe$_2$O(SeO$_3$)$_2$ taken at 290 and 6 K in five different scattering configurations.}
\end{figure*}

The temperature dependence of the magnetic susceptibility $\chi(T)$ of Fe$_2$O(SeO$_3$)$_2$ taken at $B = 0.1$ T is shown in Fig. 1(c). At high temperatures, field-cooled (FC) and zero-field-cooled (ZFC) curves coincide. The susceptibility increases with lowering temperature, culminating in a broad hump at approximately $T = 120$ K revealing the formation of short range correlations that are not described by a Curie-Weiss law [see the $1/\chi(T)$-curves in Fig. 1(d)]. Below $T_C$, the compound undergoes a magnetic phase transition into a long-range-ordered state. The low-temperature magnetic susceptibility is finite, indicating antiferromagnetism. The total, as-measured specific heat, shown in the inset of Fig. 1(d), shows a rounded peak at $T_C$, confirming a magnetic phase transition. The ZFC curve measured with the magnetic field perpendicular to the $a$ axis shows a sharp peak just below $T_C$. This may be related to a spin-flop transition in relatively low magnetic fields. A rapid increase of $\chi(T)$ below $T = 110$ K may indicate an additional ferromagnetic moment arising.

For an analysis of the coupling between lattice and spin degrees of freedom a complete knowledge of the vibrational properties of the material is essential. For the $Pccn$ orthorhombic structure of Fe$_2$O(SeO$_3$)$_2$ with three Fe ions located at the $4c$, $4d$, and $8e$ Wyckoff positions, and all selenium and oxygen ions at the $8e$ position~\cite{giester-96}, the factor group analysis gives a total of 132 Raman-active phonon modes: $\Gamma_{\mathrm{Raman}} = 32A_g + 32B_g + 34B_{2g} + 34B_{3g}$. The Raman tensors take the form:

\begin{center}

\begin{widetext} \mbox{A$_{g}$=$\begin{pmatrix} a & 0 & 0\\ 0 & b & 0\\ 0 & 0 & c\\
\end{pmatrix}$

, B$_{g}$=$\begin{pmatrix} 0 & d & 0\\ d & 0 & 0\\ 0 & 0 & 0\\ \end{pmatrix}$

, B$_{2g}$=$\begin{pmatrix} 0 & 0 & e\\ 0 & 0 & 0\\ e & 0 & 0\\ \end{pmatrix}$

, B$_{3g}$=$\begin{pmatrix} 0 & 0 & 0\\ 0 & 0 & f\\ 0 & f & 0\\ \end{pmatrix}$}

\end{widetext}
\end{center}

We find that the Raman spectra of Fe$_2$O(SeO$_3$)$_2$ at $T = 290$ K and 6 K are very well polarized [see Fig. 1(e)]. Narrow and well-distinguishable phonon peaks (accounting for a high sample quality) with various intensities are observed. In the frequency range of 10 -- 900 cm$^{-1}$ and at $T = 6$ K all 100 $\Gamma$-point Raman-active phonon modes of $A_g$, $B_{2g}$, and $B_{3g}$ symmetry can be clearly identified. In addition, very weak and broad second order excitations are observed in the higher frequency regime up to 1300 cm$^{-1}$ (not shown).

\begin{figure*}
\label{figure2}
\centering
\includegraphics[width=14cm]{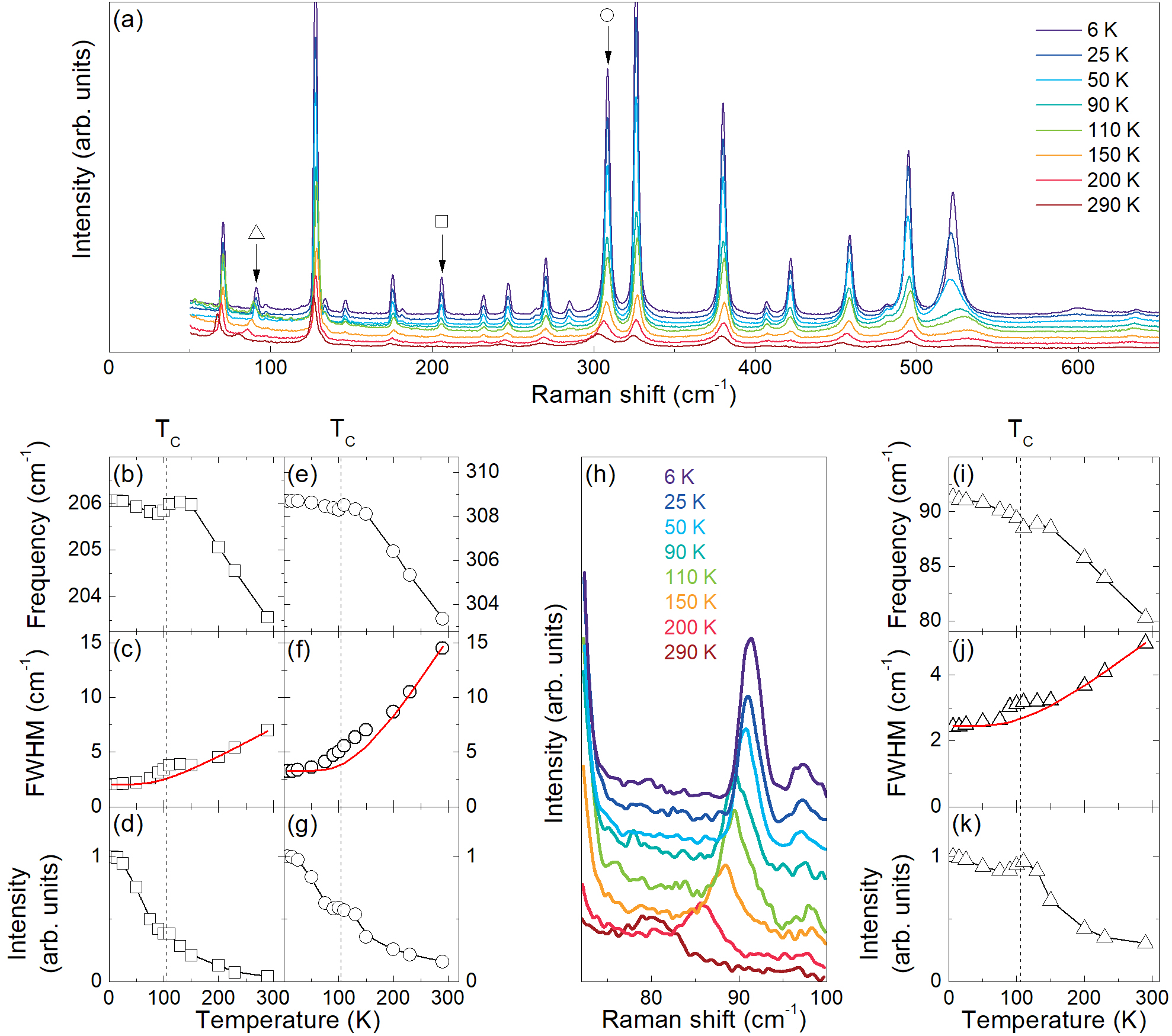}
\caption{(Color online) (a) Temperature dependent Raman spectra of Fe$_2$O(SeO$_3$)$_2$ taken in $aa$ scattering geometry. The open square, circle, and triangle mark the phonons detailed below. (b)-(g) Temperature dependence of selected phonon line parameters (frequency, width, and integrated intensity). The solid red lines represent temperature-induced anharmonic phonon behavior~\cite{wakamura-88, klemens-66, balkanski-83}. (h) Temperature dependence of the line at 80 cm$^{-1}$ (at 290 K); (i)-(k) its extracted parameters (frequency, width, and integrated intensity).}
\end{figure*}

The temperature dependence of the Raman spectra of Fe$_2$O(SeO$_3$)$_2$ obtained in $aa$ scattering geometry is shown in Fig. 2(a) and an in-depth analysis providing several distinct features is given in Figs. 2(b)-(g). First, the phonon eigenfrequencies trace the temperature dependence of the magnetic subsystem, showing a broad hump at approximately 120 K and a jump in frequency at $T_C$. This renormalization of the phonon frequency and linewidth should be ascribed to a direct contribution of an exchange coupling to the quasielastic constant of the mode, i.e., spin-phonon coupling. The deviation of the phonon linewidth behavior from an expected anharmonic broadening in the temperature region of 75 -- 150 K suggests a change of the relaxation mechanism. Furthermore, the most remarkable feature is the pronounced enhancement of phonon intensities, which grow by almost one order of magnitude upon cooling from room temperature to 6 K with a deviation from their smooth behavior in the same temperature region where linewidth anomalies appear. Similarly, an anomalous increase in the intensity of the phonon lines upon cooling was observed earlier in multiferroic compounds Cu$_2$OSeO$_3$~\cite{gnezdilov-10} and FeTe$_2$O$_5$Br~\cite{choi-14} and was unambiguously attributed to an increase of the respective electronic polarizability with temperature. Our observations indicate strong spin-lattice coupling in Fe$_2$O(SeO$_3$)$_2$.

An exception is the behavior of the low-frequency phonon mode at 80 cm$^{-1}$ (295 K). It hardens by more than 10 cm$^{-1}$ upon cooling from 290 to 6 K [see Figs. 2(h)-(i)], which is several times larger than the hardening of the remaining phonon lines. This indicates that this excitation responds especially strong to magnetic ordering.

\begin{figure*}
\label{figure3}
\centering
\includegraphics[width=18cm]{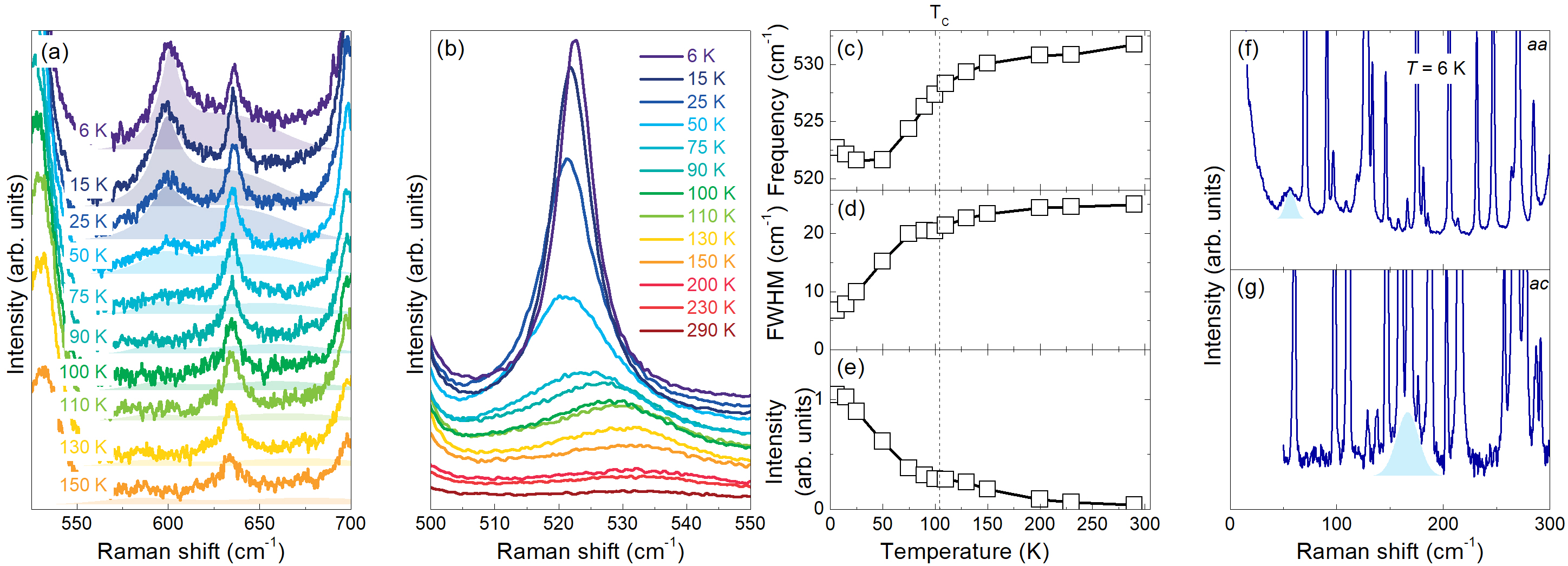}
\caption{(Color online) (a) Two-magnon Raman signal obtained in $aa$ polarization at various temperatures. Shaded areas mark the background of magnetic excitations. (b) Temperature dependence of the line at 523 cm$^{-1}$; (c)-(e) its parameters (frequency, width, and integrated intensity). (f) Low frequency magnetic excitations in the $aa$ and (g) $ac$ Raman spectra taken at $T = 6$ K.}
\end{figure*}

The most striking observation in our experiment is the set of bands appearing at low temperatures and observable only in $aa$ polarization at 54 cm$^{-1}$, 523 cm$^{-1}$ and 600 cm$^{-1}$, and a further excitation at 163 cm$^{-1}$ in $ac$ polarization. Their unique temperature dependence allows us to assign them to two-magnon Raman scattering processes related to 3D long-range order in Fe$_2$O(SeO$_3$)$_2$ (see Fig. 3).

The broad two-humped feature with a strong maximum centered at around 600 cm$^{-1}$ in the spectrum at 6 K demonstrates a typical two-magnon behavior: It softens in frequency and broadens in linewidth upon increasing temperature and strongly drops in intensity for $T > T_C$, although it remains distinguishable up to 150 K [see Fig. 3(a)]. Accordingly, short-range spin correlations within the chain survive up to $T = 1.4T_C$.

The excitation at 523 cm$^{-1}$ [Fig. 3(b)] has a comparably smaller linewidth and remains distinguishable even at room temperature. With lowering temperature its linewidth decreases in an unusual way [see Fig. 3(d)] with a local minimum at $T_C$. With further cooling, the line begins to harden below approximately 35 K [Fig. 3(c)]. Its lineshape changes with decreasing temperature from asymmetric to symmetric. Tentatively we may assign this well-defined peak to a magnetic excitation, i.e., due to singularities of the one-dimensional density of two-magnon states~\cite{konstantinovic-00}. The unusual narrow linewidth of this signal can be connected with a specific flatness of the magnetic spectrum in Fe$_2$O(SeO$_3$)$_2$. The observation of this excitation in $aa$ polarization (i.e., along the saw-tooth chain direction) is typical for spin-chain compounds and supports our tentative assignment~\cite{lemmens-03}. Similar excitations related to the magnetic dimer structure were observed, e.g., in MgV$_2$O$_5$~\cite{konstantinovic-00} and BaFe$_2$Se$_2$O~\cite{popovic-14}.

Another relatively narrow magnetic signal appears below $T_C$ at 54 cm$^{-1}$, which is one order of magnitude smaller than the energy scale of the previous magnetic signals [see Fig. 3(f)]. The signal is highly polarized and is seen only in $aa$ polarization. In addition, a signal seen at 163 cm$^{-1}$ in crossed polarization [see Fig. 3(g)] also has a relatively narrow linewidth. Its observation in crossed $ac$ polarization reflects the specific structure of the saw-tooth units with comparable distances between magnetic ions along $a$ and $c$ directions. In crossed polarization magnetic excitations of different symmetry can contribute to the Raman signal. Therefore, its position represents some new energy scale existing in the magnetic spectrum of Fe$_2$O(SeO$_3$)$_2$. We summarize all magnetic modes and their properties in Table I.

\begin{table*} \caption{\label{tab:table1}Magnetic excitations in Fe$_2$O(SeO$_3$)$_2$.}
\begin{ruledtabular} \begin{tabular}{ c|c|c|c|c }
 $Frequency$&$Width$&$Pol.$&$Properties$&$Tentative$\\
 ($T = 6$ K, cm$^{-1}$)&($T = 6$ K, HWHM, cm$^{-1}$)& & &$assignment$\\ \hline
  & & & &2 formerly\\
 54.7&12&$aa$&Chain specific pol.&flat band zero\\
  & & & &energy magnons\\ \hline
  & & & &High energy +\\
  & & & &one of the\\
 166.6&25&$ac$&Crossed pol.&activated zero\\
  & & & &energy flat\\
	& & & &band magnons\\ \hline
  & & & &High energy +\\
	& & &visible up to $T = 295$ K, linewidth&one of the\\
 522.7&7&$aa$&twice as large as the phonon lines,&activated zero\\
  & & &chain specific pol.&energy flat\\
	& & & &band magnons\\ \hline
  & & &Two components, 601 cm$^{-1}$ and&High energy +\\
  & & &634.5 cm$^{-1}$, broad maximum,&one of the\\
 620&$<$100&$aa$&visible up to 150 K, anomalous&activated zero\\
	& & &frequency, width, and intensity,&energy flat\\
	& & &chain specific pol.&band magnons\\
\end{tabular} \end{ruledtabular} \end{table*}

The observed signals share some general feature -- their relatively narrow linewidth compared to the scale of exchange that follows from the rather high $T_C$. Since there are sixteen magnetic Fe ions per unit cell, the spin-wave spectrum of Fe$_2$O(SeO$_3$)$_2$ contains sixteen branches [see Figs. 1(a) and 1(b)]. The unusual diversity of two-magnon bands by energy evidences a strict separation of the spin-wave spectrum into low and high energy spin waves without any overlapping in between. The narrow linewidth of the two-magnon bands indicates the existence of flat regions in the magnetic spectrum. All of this is a hallmark of the saw-tooth lattice and a magnetic structure where low energy excitations originate from formerly zero energy flat-band modes. In the following we will discuss these features in detail.

\section{Discussion}

We will first focus on structural details of Fe$_2$O(SeO$_3$)$_2$ which affect the aforementioned properties. The Fe ions build up the four saw-tooth chains that are running along the $a$ axis. The spin chain is formed by a sequence of corner-sharing Fe1O$_6$ octahedra and Fe2O$_4$ tetrahedra. The tilted teeth of the chain are formed by Fe3O$_6$ oxygen octahedra which are connected to Fe1O$_6$ octahedra by edge-sharing. The saw-tooth chains connect to each other along the $b$ axis by edge-sharing Fe3O$_6$ octahedra. Along the $c$ axis the saw-tooth chains are connected via (SeO$_3$)$^{2-}$ complexes changing the Fe-O-Fe exchange pathway into a longer Fe-O-Se-O-Fe one, as sketched in Fig. 4(a).

\begin{figure}
\label{figure4}
\centering
\includegraphics[width=8cm]{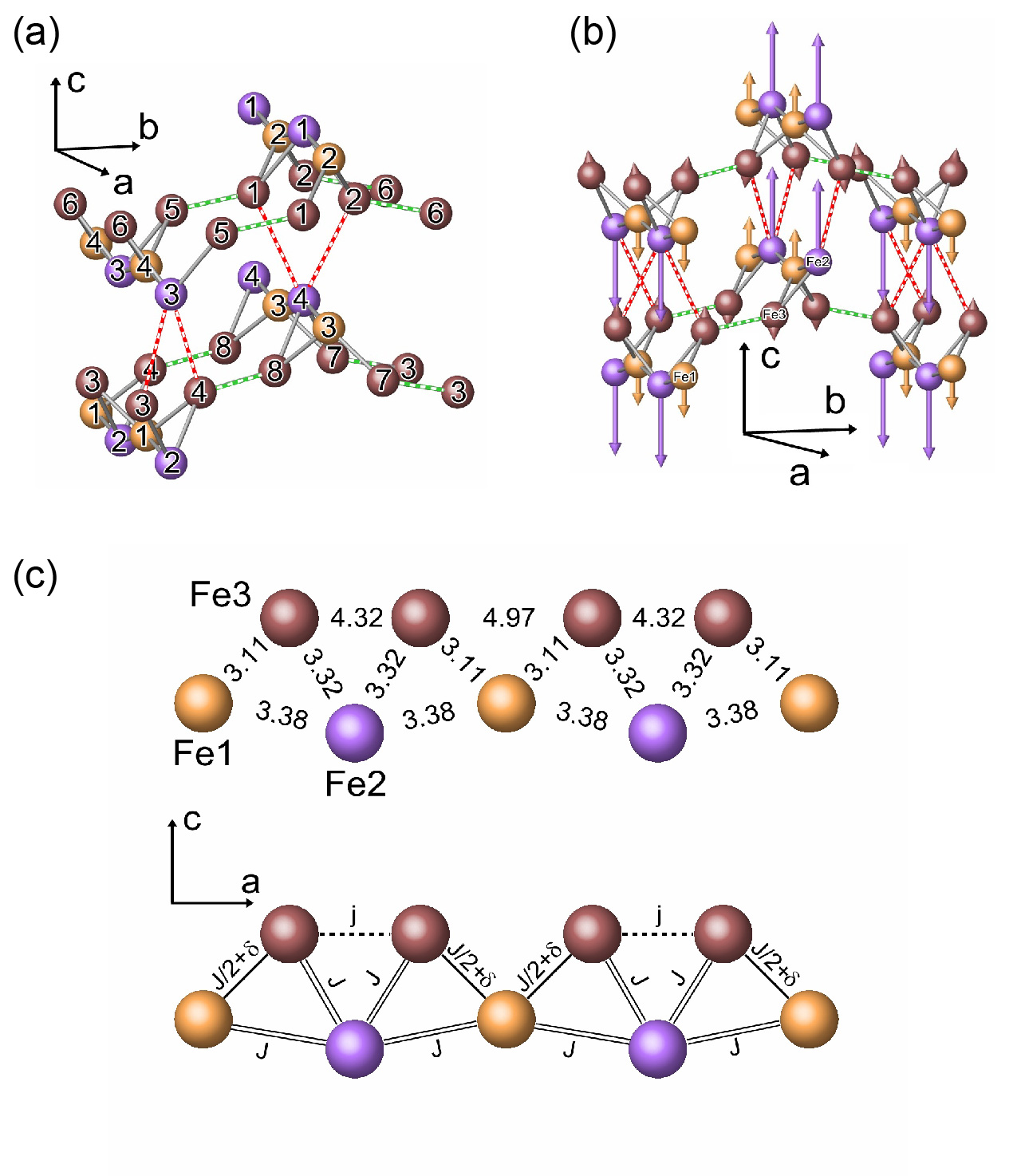}
\caption{(Color online) (a) Schematics of the magnetic Fe$^{3+}$ ion positions in Fe$_2$O(SeO$_3$)$_2$. All oxygen and selenium ions are omitted for clarity. The Fe ions in $4c$, $4d$ and $8e$ Wyckoff sites are shown in gold, violet and dark brown respectively. The intralayer interchain exchanges are marked by dashed green lines. The interlayer interactions are realized through Fe2-O-Se-O-Fe3 bonds and highlighted by dashed red lines. The accepted enumeration of Fe ions is shown. (b) Vibrational pattern of the low-frequency $A_g$ mode experimentally observed at frequencies of 80-92 cm$^{-1}$. For clarity only Fe$^{3+}$ ions are shown. The length of arrows represents relative amplitudes of out-of-phase motion of the Fe2 and Fe3 ions from the neighboring Fe$^{3+}$ layers. (c) Mapping of the structure on a saw-tooth magnetic unit. The oxygen and selenium ions are omitted for clarity. A possible approximation of exchange constants is shown. Note that the saw-tooth chain includes two different teeth instead single tooth in the usual saw-tooth chain. This double-tooth chain requests four magnetic ions for a minimal model.}
\end{figure}

Looking at the arrangement of magnetic bonds (e.g. in the approximation of nearest-neighbor exchange interactions) one can sort out two weakly coupled Fe$^{3+}$ $ab$ planes which are formed by two saw-tooth chains running along the $a$ axis. The interaction between these chains is weaker than the intrachain interactions. Indeed, the angle of the Fe1-O-Fe2 bonds in corner-shared octahedra along the spine of chains is larger than the angle of the Fe3-O-Fe3 bonds in edge-shared octahedra connecting neighboring saw-tooth chains. In contrast, the corner-shared Fe2-O-Fe3 intrachain interaction between vertex and spine Fe ions remains comparable with Fe1-O-Fe2 interactions. This hierarchy supposes that the 1D saw-tooth chain is the main magnetic unit in Fe$_2$O(SeO$_3$)$_2$, and the interplane Fe2-O-Se-O-Fe3 exchange interaction should be responsible for the 3D long-range ordering.

\begin{figure}
\label{figure5}
\centering
\includegraphics[width=8cm]{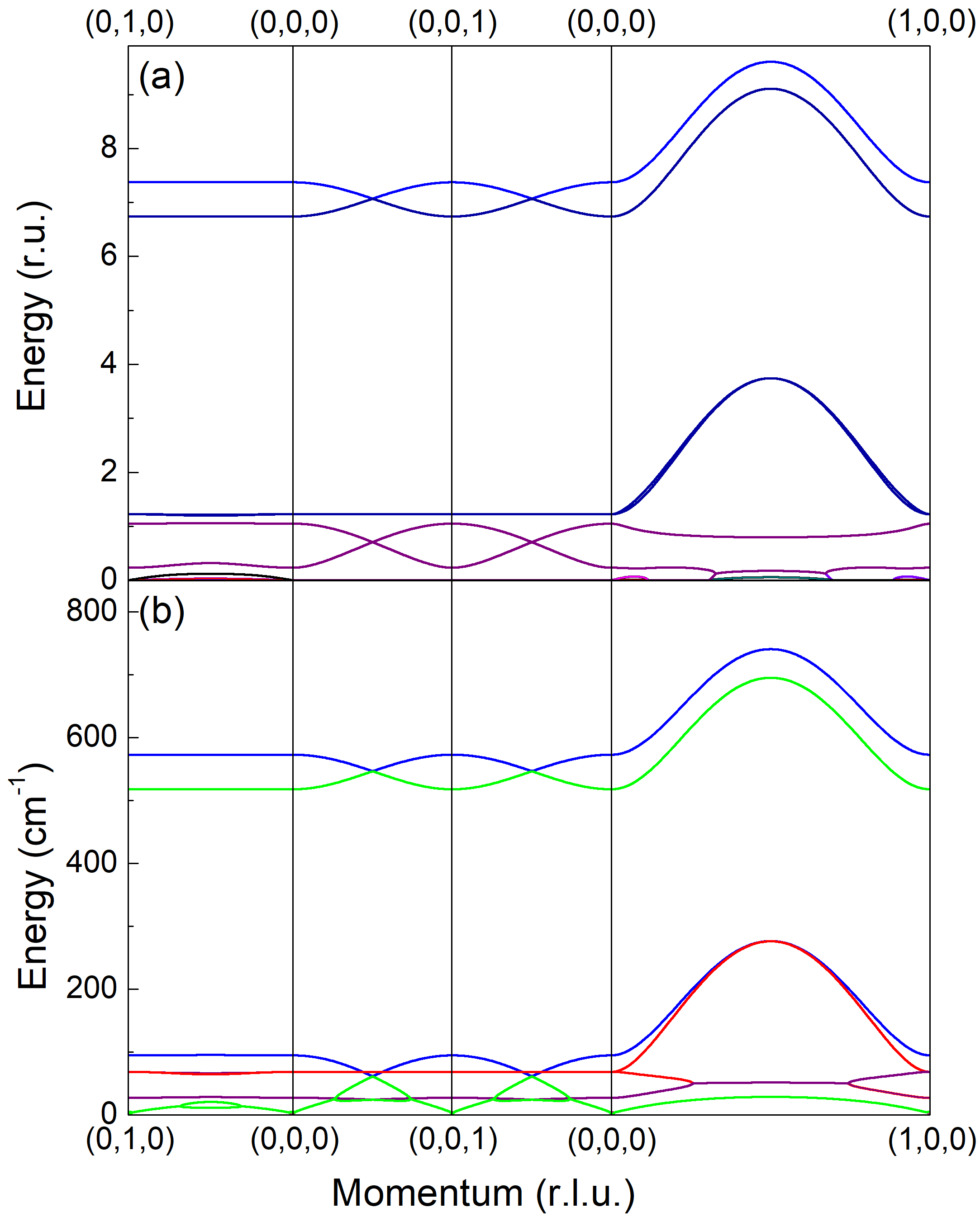}
\caption{(Color online) (a) One-magnon spectrum along high symmetry lines of four coupled saw-tooth chains mapped to the minimal model in Fig. 4(c). The Fe1-Fe2 exchange $J$ is selected as the energy unit, and the AFM intralayer exchange $I_l = 0.1J$, the interlayer AFM exchange $I_p = 0.1J$, the vertex-vertex (Fe3-Fe3) AFM exchange $j = 0.1J$, and $\delta = 0$. The development of the incommensurable instability along the $a$ axis is seen as the appearance of zero frequencies at a certain wave vector along (100). This wave vector will be the helix wave vector. (b) The calculated one-magnon spectrum for exchange parameters estimated from two-magnon signals. The blue and green colors denote double degenerate magnon branches which have the same symmetry at $k=0$. The double degenerate Goldstone modes (in green color) merge with the two lowest gapped branches with the gap at 5 cm$^{-1}$. The red color denote the four degenerate and almost flat-band modes along the (010) and (001) directions. Similar double-degenerate flat-band mode is seen at 27 cm$^{-1}$.}
\end{figure}

Our Raman data [see Figs. 2(h)-(k)] and lattice dynamic calculations [see Fig. 4(b)] strongly support the assumption, that the phonon mode corresponding to out-of-phase motions of the Fe2 and two Fe3 ions from neighboring Fe$^{3+}$ $ab$ planes should be highly sensitive to 3D magnetic order. As we have shown by our lattice dynamical calculations, one of the lowest-frequency Raman modes of $A_g$ symmetry (at 80 cm$^{-1}$) represents the displacements of Fe2 ions along the $z$ axis. In neighboring (along $y$ direction) chains these displacements are anti-phase (for details, see Supplement). In such a type of motion, the distance varies along the $z$ direction between neighboring Fe2 and Fe3 ions. Their exchange coupling provides the three-dimensional magnetic ordering and leads to a renormalization of this phonon mode.

The susceptibility measurements suggest that below $T_C$ the magnetic long range order is antiferromagnetic [see Fig. 1(c)-(d)], and the easy axis (i.e., main antiferromagnetic vector) should be perpendicular to the $a$ axis. Measurements perpendicular to the $a$ axis show that the applied magnetic field has components both perpendicular and parallel to this easy axis. An anomalous increase of the susceptibility just below $T_C$ as well as the large difference for FC and ZFC data can be interpreted as a presence of a weak ferromagnetic moment which points perpendicular to the $a$ axis. However, the development of an incommensurate magnetic structure can also be a cause of the susceptibility anomalies. An increase of electronic polarizability below $T_C$, which is detected in our Raman spectra, points to such a possibility. Only an incommensurate magnetic order can be a source of electronic polarization in the crystal with space inversion symmetry~\cite{harris-06}.

Fig. 4(c) illustrates the saw-tooth chain unit including the distances between Fe ions. The mutual relations of exchange interaction constants can be roughly estimated based on the type and angles of the Fe-O-Fe bonding. Note that Fe1-Fe2 and Fe2-Fe3 ions are connected through corner-shared octahedra while Fe1-Fe3 ions are connected through edge-shared octahedra. The Fe3-Fe3 exchange is realized through a Fe-O-Se-O-Fe pathway.

Here we assume that Fe1-Fe2 (with an angle of Fe1-O-Fe2 = 127.79$^{\circ}$) and Fe2-Fe3 (with an angle of Fe2-O-Fe3 = 123.7$^{\circ}$) are bonded by the same AFM exchange. The Fe1-Fe3 bonding is weaker, as it forms through Fe1-O2-Fe3 plaquettes of edge-shared Fe1 and Fe3 octahedra with both angles Fe-O-Fe (Fe-O'-Fe = 93.24$^{\circ}$ and Fe-O''-Fe = 107.95$^{\circ}$) being closer to the critical value where the exchange interaction changes sign. The Fe3-O-Se-O-Fe3 exchange $j$ between the tooth's vertices with a distance of 4.32 {\AA} is supposed to be nonzero as it has the same origin as the interlayer Fe2-O-Se-O-Fe3 exchange with a distance of 4.64 {\AA}. It's worth repeating that this last interaction is responsible for establishing long-range order.

To elucidate the general structure of the magnon spectrum and to check the stability of uniform magnetic order we proceed with the one-magnon spectra calculation with small variations of the $\delta$ and $j$ constants compared to the exchange $J$ and supposing that the interlayer exchange $I_p$ and intralayer exchange $I_l$ can be either FM or AFM. A detailed analysis of possible types of uniform exchange magnetic long-range order in Fe$_2$O(SeO$_3$)$_2$ is presented in the Supplement. The calculations have been performed in linear spin wave (LSW) approximation with the help of the SpinW code~\cite{toth-15}. The LSW theory can capture an incommensurate instability of the given 3D uniform long-range order under detection of some spin wave approaching zero energy at a certain wave vector. An example of a spin wave spectrum with an incommensurate instability of the initial collinear phase with both $I_p$ and $I_l$ being AFM ($\Gamma_4$ phase) is shown in Fig. 5(a).

The respective numerical analysis has been done for all four uniform structures which can be realized for different signs of interlayer ($I_p$) and intralayer ($I_l$) interchain exchanges (see Supplement for details). We conclude that the structure with an AFM intralayer exchange $I_l > 0$ and AFM interlayer exchange $I_p > 0$ supports the development of an incommensurate magnetic order. A uniform exchange structure with ferromagnetic interlayer exchange $I_p < 0$ and $I_l > 0$ ($\Gamma_2$ phase) can be stable if the modulus of $|I_p|$ is above a certain threshold which is roughly one order smaller than the Fe1-Fe3 intrachain interaction. The uniform exchange magnetic structures with the ferromagnetic intralayer interchain exchange $I_l < 0$ are always stable, irrespective of the sign of the interlayer exchange ($\Gamma_1$ phase and $\Gamma_3$ phase). Thus, in the following we will consider as a starting point the $\Gamma_4$ phase, with all exchanges positive, as a classical ground state of the system.

The above mapping shows a distinct difference in the topology of the exchange patterns of the saw-tooth chain in Fe$_2$O(SeO$_3$)$_2$ compared to a standard saw-tooth chain ($\Delta$ chain) with equal values of the exchange interactions between spine and vertex (e.g. for the case $\delta=J/2$ and $j=0$)~\cite{kubo-93, nakamura-96, sen-96}. One may check the existence of zero-energy flat-band modes in the classical limit for the isolated, non-interacting saw-tooth unit in Fe$_2$O(SeO$_3$)$_2$ [see Fig. 4(c)]. Most unusual is the presence of double degenerate zero-energy flat-band modes in the case of $j = 0$ and $\delta = 0$. This feature is related to four magnetic ions in the primitive cell of the saw-tooth unit instead of two ions in the $\Delta$ chain. In the case of $j \neq 0$ and $\delta = 0$ one flat-band mode survives and one Goldstone mode appears. Two high energy modes emerge with energies of $2.5SJ$ and $S(J/2-2j)$ at the $\Gamma$ point. No flat-band modes exist for $\delta \neq 0$ independently of the value of $j$. Note that the presence of the vertex-vertex interaction $j$ does not contribute to a gap formation. Furthermore, the zero-energy flat-band modes disappear in the case of different values of $J_{bb}$ (spine-spine, Fe1-Fe2) and $J_{bv}$ (spine-vertexes, Fe2-Fe3) exchange. The activated flat-band modes remain weakly dispersive with a gap dependent on $\delta$ and the difference $J_{bb} - J_{bv}$.

In the classical limit the spin wave spectrum of interacting saw-tooth units in Fe$_2$O(SeO$_3$)$_2$ with $I_l$, $I_p \ll J$ is separated into several energy regions. It is dominated by the presence of a few high-energy SW branches and a large number of low-energy SW branches, which partially originate from formerly zero-energy flat-band modes. In contrast to conventional 3D antiferromagnets the spin waves from different energy regions do not overlap in $k$-space.

The flat-band modes survive for the 2D and 3D structures of interacting saw-tooth chains with $\delta = 0$. Note that for non-interacting four saw-tooth units in the crystallographic unit cell of Fe$_2$O(SeO$_3$)$_2$ every mode will be fourfold degenerate. In the 2D $ab$ layer of saw-tooth chains interacting through the interchain exchange $I_l$ the zero-energy flat mode survives along the (010) direction. In the 2D $ac$ layer of saw-tooth chains connected through $I_p$ exchange along the (001) direction, the double-degenerate zero-energy flat-band mode exists in the (001) direction. These modes survive along the (001) direction even in the 3D case with both non-zero $I_p$ and $I_l$ exchanges, whereas zero-energy flat-band modes along the (010) direction are absent. The respective model spectrum is shown in Fig. 5(a).

The value of $\delta$ defines the activation energy of the formerly zero-energy flat-band modes. We find the four-fold degenerate SW at an energy of $S\sqrt{E(E+2I_l)}$ at $k = 0$, where $E = J/2-2j-\delta+I_p$ remains almost non-dispersive along the (010) and (001) directions under arbitrary values of 3D interchain couplings. The energetically highest four-fold degenerate modes split into two double degenerate modes under consideration of interlayer interchain coupling $I_p$. We reveal that the splitting linearly depends on the value $I_p$. The dispersion of high-energy modes in the Heisenberg approximation does not show any obvious sensitivity to the underlying incommensurate order.

The observed two-magnon Raman bands contain an unusual combination of high- and low-energy magnon branches. The necessary symmetry conditions for two-magnon scattering in parallel $aa$ polarization are fulfilled to obey such a contribution. Indeed, every four (out of sixteen) magnon branches have the same symmetry along (100) (010) and (001) directions in the Brillouin zone, and every set of branches with the same symmetry includes at least one high-energy branch and at least two low-energy branches. The specific saw-tooth structure allows two-magnon scattering in the non-diagonal $ac$ polarization with the participation of two magnon branches of different symmetry. The quasi 1D exchange topology in Fe$_2$O(SeO$_3$)$_2$ implies a very weak magnon dispersion perpendicular to the spine of chains that forms a high density of states for spin waves with energies close to the $\Gamma$ point. We conclude that the energy scale of the two-magnon bands in Fe$_2$O(SeO$_3$)$_2$ observed in our Raman data is given by magnon energies at the $\Gamma$ point and not at the Brillouin zone boundaries as is realized in conventional 3D antiferromagnets. The anomalous narrow linewidth of the two-magnon bands, comparable with the phonon linewidth, can be explained by the flatness of spin waves contributing to the scattering process. 

Energies of the spin waves at the $\Gamma$ point have been calculated analytically in the nearest neighbor Heisenberg approximation (see Supplement). Using this approach, we estimate the exchange constants of Fe$_2$O(SeO$_3$)$_2$: $J = 75.84$ cm$^{-1}$; $I_p = 9.18$ cm$^{-1}$; $j = 12.1$ cm$^{-1}$; $I_l = 10$ cm$^{-1}$ and $\delta = 3.77$ cm$^{-1}$. Based on these exchange energies we estimate the Neel temperature $T_N = 103 - 106$ K, i.e., fully coinciding with the experiment. The restored spin wave spectrum of Fe$_2$O(SeO$_3$)$_2$ is shown in Fig. 5(b). The spectrum does not show any incommensurate instability for the given choice of parameters. However, a small variation of the interlayer interchain exchange $I_l$ and a respective change of the ratio $j/I_p$ both introduce incommensurability along the (100) direction. There are two flat-band modes almost dispersionless on $k_yk_z$-plane of $k$ space at 68.4 cm$^{-1}$ and 27 cm$^{-1}$ present in the spectrum. A very weak dispersion for these modes is observed along the (010) direction.

These interaction constants with antiferromagnetic vertex-vertex interaction highlight the strong degree of frustration existing in Fe$_2$O(SeO$_3$)$_2$. At the same time they may provide an explanation of the instability of the fluctuating quantum dimer state. Indeed, a simple analysis of the experimental magnetic susceptibility shows that in the range between $T_C$ and 300 K the magnitude of $\chi_{\mathrm{exp}}(T)$ lies far below $\chi_{\mathrm{CW}}(T)$ [see Fig. 1(c)], which accounts for the Curie-Weiss susceptibility of sixteen unbound Fe$^{3+}$ ions (eight formula units per primitive cell). This discrepancy can be unambiguously attributed to the presence of various dynamic valence bonds in their singlet ground states and with a high enough energy of its excited magnetic levels. Therefore, they do not contribute to the magnetic susceptibility. However, in spite of the relatively low values of the interchain interactions which validate the 1D approximation, we found that the ratio of spine-vertex (Fe1-Fe3) to spine-spine (Fe1-Fe2) interactions $J_{bv}/J_{bb} = (J/2+\delta)/J$ is close to 1/2. As pointed out in Ref.~\cite{blundell-03}, the difference in $J_{bb}$ and $J_{bv}$ interactions leads to a suppression of the spin gap in the quantum limit for the $S = 1/2$ saw-tooth model chain with a dimer ground state. The spin gap approaches zero at $k = 0$ for $J_{bb}/J_{bv}$ close to 0.5 and at $k = \pi$ for $J_{bb}/J_{bv}$ close to 1.5. The previously dispersionless $S = 1$ excitations become highly dispersive and the system turns into Heisenberg chains.

Taking DM interaction into account should have a similar influence, even if $J_{bb} = J_{bv}$. As shown in Ref.~\cite{hao-11}, a relatively weak DMI of $\sim 0.1 J_{bb}$ is sufficient to destroy the valence-bond order, to close the spin gap, and to turn the system into a Luttinger liquid with incommensurate spin correlations and spin-wave excitations. In Fe$_2$O(SeO$_3$)$_2$ DMI contributes to all Fe-Fe magnetic bonds. As follows from bond symmetry the DM vector for the Fe1-F2 spine-spine magnetic bonds is directed purely along the $z$ axis, whereas for all other bonds it has all Cartesian components. To this point one can add that DMI also supports the onset of an incommensurable instability of a uniform 3D long range order. In the classical limit, the presence of DM interaction forms a spiral order in the usual $S = 1/2$ saw-tooth chain~\cite{hao-11}. Note that DMIs were not included in the above considerations as we show by model calculations that the incommensurate instability in Fe$_2$O(SeO$_3$)$_2$ can be induced already in the Heisenberg exchange approximation.

\section{Summary}

In the saw-tooth spin-chain system Fe$_2$O(SeO$_3$)$_2$ long-range magnetic order is established below $T_C = 105$ K, as confirmed by anomalies in specific heat and magnetic susceptibility. Raman-active $A_g$ phonon modes show anomalies in linewidths and energies close to $T_C$. The $A_g$ mode at 80 cm$^{-1}$, which constitutes the out-of-phase motion of the Fe ions from neighboring Fe layers, is especially sensitive to the onset of 3D order. An anomalous increase of phonon intensities below $T_C$ points towards an increase of the electronic polarizability. We conclude an incommensurability of the magnetic 3D order in Fe$_2$O(SeO$_3$)$_2$ which would give rise to electronic polarization in a crystal with space inversion symmetry.

Excitations related to long-range magnetic order exist with polarization along the chain direction with a pronounced temperature dependence. We assign them to two-magnon signals. These excitations are strongly differentiated in energy with relatively small linewidth in contrast to a widely spread continuum which is expected from the high $T_C$ of this compound. We connect these features to the specific form of the one-magnon spectrum of the saw-tooth magnetic lattice with a set of dispersionless spin-wave branches. The minimal set of exchange coupling constants has been roughly estimated based on Raman data and results of linear spin-wave calculations. The onset of the incommensurate 3D magnetic order and the vanishing of the dimer quantum state in the highly frustrated title compound may result from the specific relations between given exchange interactions. We show that in the classical limit the one-magnon spectrum of Fe$_2$O(SeO$_3$)$_2$ contains a few almost flat-band modes in the $k_yk_z$ plane of $k$ space.

\begin{acknowledgments}
This work has been supported by the Russian Ministry of Education and Science of the Russian Federation through NUST ``MISiS'' Grant No. K2-2017-084 and by the Act 211 of the Government of Russia, Contracts 02.A03.21.0004 and 02.A03.21.0011. We also want to thank RFBR project 16-03-00463a for the financial support, the ``Nieders\"{a}chsisches Vorab'' through ``Quantum- and Nano-Metrology (QUANOMET)'' initiative within the project NL-4, and the DFG-LE967/16-1 Project.
\end{acknowledgments}

\section{Supplementary}

\subsection{Crystal Growth}

Single crystals of Fe$_2$O(SeO$_3$)$_2$ were grown by the chemical vapor transport (CVT) technique. Fe$_2$O$_3$, FeCl$_3 \cdot$ 6H$_2$O, and SeO$_2$ were mixed in the molar ratio 0.8 : 0.2 : 4 (about 1 g of total mixture weight) inside a dry argon filled glove box and loaded into a silica tube with $\sim 18$ mm in diameter and 12.5 cm length. The tube was sealed under vacuum (about 50 mTorr) and loaded into a two-zone furnace. Both zones were heated to 300 $^{\circ}$C for 24 h. Subsequently, the temperature in both zones was raised to 400 $^{\circ}$C within 12 h. At the last step, the evaporation zone containing the starting mixture was heated up to 480 $^{\circ}$C and the temperature gradient 480 $^{\circ}$C / 400 $^{\circ}$C was kept for 10 days. After cooling to room temperature, large needle-like dark brown crystals with up to 1 cm length were obtained in the deposition zone of the reaction tube. XRD experiments performed on a STOE IPDS diffractometer confirmed the resulting crystals as Fe$_2$O(SeO$_3$)$_2$. The unit cell constants are in a good agreement with those reported previously by Giester~\cite{giester-96}.

\subsection{Lattice Dynamical Calculations}

In order to assign the symmetries and eigenvectors of the observed optical phonon modes, we compute the $\Gamma$-point phonon modes by adopting shell-model lattice dynamical calculations implemented in the General Utility Lattice Program (GULP) package~\cite{gulp}. Within the shell model, we treat Fe and Se ions as point cores with charges $X$ and O ions as a sum of point cores and a massless shell with charge $Y$, representing valence electrons. The interionic interactions between Fe/Se and O ions are described by short-range Born-Mayer-Buckingham potentials between ions $i$ and $j$:

$V_{\mathrm{BM}}(r) = A_{ij} \mathrm{exp}(-r/\rho_{ij}) - C_{ij}/r^6$,
where $A_{ij}$ and $\rho_{ij}$ denote the strength and the range of the repulsive interaction, respectively, and $C_{ij}$ describes an attractive part with the interatomic distance $r$. See Table II for the corresponding values.

\begin{table*} \caption{\label{tab:table2}Values of the shell model parameters for Born-Mayer-Buckingham potentials used in our calculations.}
\begin{ruledtabular} \begin{tabular}{ c|c|c|c|c|c|c }
 Atom&$X(\vert e \vert)$&$Y(\vert e \vert)$&Atomic pair&A (eV)&$\rho$ (eV)&C (eV \AA$^6$)\\ \hline
 Fe&+3&0&Fe-O&1414.6&0.3128&0\\
 Se&+4&0&Se-O&3414.6&0.3128&0\\
 O&0.86902&-2.86902&O-O&22764.3&0.149&27.88\\
\end{tabular} \end{ruledtabular} \end{table*}

In Table III we compare experimental and calculated frequencies of A$_g$ phonon modes. A remarkably good agreement is found, in spite of the adopted simple potentials and challenges in lattice dynamical calculations associated with the complexity of the crystal structure and the lack of the interatomic potential parameters for Se in existing databases.

\begin{table*} \caption{\label{tab:table3}Experimental (at $T = 6$ K) and calculated phonon frequencies in cm$^{-1}$ of fundamental A$_g$ modes. The contributions of magnetic ions to each mode are given. The normalized eigenvectors are shown for a representative ion from each $4c$ (Fe1), $4d$ (Fe2), and $8e$ (Fe3) Wyckoff position [cf. Fig. 4(a), main text]. The contributions of magnetic ions to high frequency phonon modes are almost negligible.}
\begin{ruledtabular} \begin{tabular}{ c|c|c|c|c|c|c||c|c|c|c|c|c|c }
  & &\multicolumn{5}{c||}{Magnetic ions contribution}& & &\multicolumn{5}{c}{Magnetic ions contribution}\\ 
	\cline{3-7} \cline{10-14}
 Experiment&Theory&Fe1&Fe2&\multicolumn{3}{c||}{Fe3}&Experiment&Theory&Fe1&Fe2&\multicolumn{3}{c}{Fe3}\\
	& &$z$&$z$&$x$&$y$&$z$& & &$z$&$z$&$x$&$y$&$z$\\ \hline
 92.5&88.5&-0.15&0.30&0.02&-0.01&-0.09&460.4&463.1&0.03&-0.12&0.07&0.14&0.07\\
 129.3&134.5&-0.07&-0.05&-0.03&0.01&0.07&482.6&485.9&--&-0.03&-0.05&-0.02&-0.01\\
 152.1&154.8&0.19&-0.09&-0.04&0.0&-0.10&502.3&507.7&0.02&-0.03&0.04&-0.10&0.10\\
 177.2&172.1&0.06&-0.08&0.13&-0.07&-0.22&525.0&532.7&--&--&-0.08&-0.06&0.0\\
 207.6&210.4&-0.26&-0.11&0.02&0.0&-0.03&537.4&542.1&0.05&0.02&0.01&0.11&-0.01\\
 233.5&239.7&0.10&-0.02&0.14&0.07&0.03&558.6&568.0&0.09&-0.07&-0.08&0.01&0.02\\
 266.0&266.7&0.08&0.06&-0.10&0.03&0.06&572.2&571.5&0.09&--&0.08&0.10&0.0\\
 272.0&278.1&0.09&--&0.04&-0.19&0.06&624.2&618.7&--&-0.06&0.07&-0.08&0.04\\
 277.5&279.8&-0.06&0.11&0.02&0.03&0.0&635.8&625.8&0.05&--&-0.06&-0.02&-0.02\\
 286.4&297.5&-0.13&-0.16&--&--&--&666.9&665.0&0.07&-0.07&-0.01&-0.03&0.01\\
 310.4&307.1&0.08&0.08&0.03&0.03&0.11&720.2&738.6&0.10&0.04&-0.03&-0.04&-0.01\\
 327.8&325.5&-0.05&-0.06&-0.07&0.04&0.0&748.2&756.4&0.05&-0.02&--&--&--\\
 381.5&378.1&0.02&0.09&0.11&0.02&0.02&776.1&787.8&0.05&--&0.02&0.04&-0.01\\
 386.5&387.4&0.03&0.13&0.02&-0.05&0.08&799.7&807.8&--&0.05&--&--&--\\
 408.7&403.8&0.13&0.10&-0.07&0.02&-0.07&843.0&853.3&--&--&0.09&-0.02&0.01\\
 424.0&429.2&-0.05&-0.09&-0.05&-0.02&0.01&856.8&856.6&--&--&0.07&-0.08&0.06\\
\end{tabular} \end{ruledtabular} \end{table*}

\subsection{Symmetry Analysis of Magnetic Degrees of Freedom and Spin Wave Spectrum}

To analyze the long range magnetic order that can be realized in Fe$_2$O(SeO$_3$)$_2$ we introduce linear combinations of the Fourier-transformed spins $\vec{s}_{\alpha}^{(i)} = \vec{s}_{\alpha}^{(i)}(\vec{k}) = \sum_{n} = \vec{s}_{\alpha n}^{(i)} \mathrm{e}^{-i\vec{k}\vec{R}_n}$ of the Fe($i$) sublattices. Here $i = 1,2,3$ enumerates the $4c$ (Fe1), $4d$ (Fe2) and $8e$ (Fe3) Wyckoff sites and the index $\alpha$ is the number of the associated Fe ion at the respective sites [for enumeration see Fig. 4(a)].

\begin{equation}
\label{eq:1}
\begin{aligned}
\vec{F}^{(i)} = &\vec{s}_1^{(i)}+\vec{s}_2^{(i)}+\vec{s}_3^{(i)}+\vec{s}_4^{(i)}; \\
\vec{L}_1^{(i)} = &\vec{s}_1^{(i)}+\vec{s}_2^{(i)}-\vec{s}_3^{(i)}-\vec{s}_4^{(i)}; \\
\vec{L}_2^{(i)} = &\vec{s}_1^{(i)}-\vec{s}_2^{(i)}+\vec{s}_3^{(i)}-\vec{s}_4^{(i)}; \\
\vec{L}_3^{(i)} = &\vec{s}_1^{(i)}-\vec{s}_2^{(i)}-\vec{s}_3^{(i)}+\vec{s}_4^{(i)}; \\
\vec{F}^{(\pm)} = &\vec{s}_1^{(3)}+\vec{s}_2^{(3)}+\vec{s}_3^{(3)}+\vec{s}_4^{(3)} \\
            &\pm (\vec{s}_5^{(3)}+\vec{s}_6^{(3)}+\vec{s}_7^{(3)}+\vec{s}_8^{(3)}); \\
\vec{L}_1^{(\pm)} = &\vec{s}_1^{(3)}+\vec{s}_2^{(3)}-\vec{s}_3^{(3)}-\vec{s}_4^{(3)} \\
              &\pm (\vec{s}_5^{(3)}+\vec{s}_6^{(3)}-\vec{s}_7^{(3)}-\vec{s}_8^{(3)}); \\
\vec{L}_2^{(\pm)} = &\vec{s}_1^{(3)}-\vec{s}_2^{(3)}+\vec{s}_3^{(3)}-\vec{s}_4^{(3)} \\
              &\pm (\vec{s}_5^{(3)}-\vec{s}_6^{(3)}+\vec{s}_7^{(3)}-\vec{s}_8^{(3)}); \\
\vec{L}_3^{(\pm)} = &\vec{s}_1^{(3)}-\vec{s}_2^{(3)}-\vec{s}_3^{(3)}+\vec{s}_4^{(3)} \\
              &\pm (\vec{s}_5^{(3)}-\vec{s}_6^{(3)}-\vec{s}_7^{(3)}+\vec{s}_8^{(3)}); \\
\end{aligned}
\end{equation}

We start from the assumption that 3D long-range magnetic order in Fe$_2$O(SeO$_3$)$_2$ can occur without any multiplication of the crystallographic unit cell. Corresponding magnetic order parameters are given by their Cartesian components of combinations in Eqn. (1) which should be taken at $\vec{k} = 0$. These magnetic order parameters (or magnetic modes) transform in accordance with their irreducible representations (IRP) at the $\Gamma$-point of the space group $Pccn$ (\# 56) (see Table IV). Here $F$ represents the ferromagnetic moment and $L$ the antiferromagnetic vectors. In the continual limit these quantities represent the magnetization and the staggered magnetization, respectively. In this limit only rotational elements of the space group remain important.

\begin{table*} \caption{\label{tab:table4}Magnetic modes of the magnetic propagation vector $\vec{k} = 0$, i.e., basis functions of the irreducible representations $\Gamma_i (i = 1-8)$. The $x,y,z$ indices indicate the non-zero Cartesian components of the vectors [Eqn. (1)]. The Cartesian components of the magnetic ($\vec{m}$) and electric ($\vec{p}$) moments are given in the left column.}
\begin{ruledtabular} \begin{tabular}{ c|c|c|c|c|c|c|c|c|c|c|c }
  &$e$&$2z$&$2y$&$2x$&$I$&$m_z$&$m_y$&$m_x$&Fe1&Fe2&Fe3\\ \hline
 $\Gamma_1$&+&+&+&+&+&+&+&+&$L_{2z}^{(1)}$&$L_{2z}^{(2)}$&$L_{3x}^{(+)}, L_{2y}^{(+)}, L_{1z}^{(+)}$\\
 $\Gamma_2$&+&+&+&+&-&-&-&-&$L_{3z}^{(1)}$&$L_{3z}^{(2)}$&$L_{3x}^{(-)}, L_{2y}^{(-)}, L_{1z}^{(-)}$\\
 $\Gamma_3 m_z$&+&+&-&-&+&+&-&-&$F_{z}^{(1)}$&$F_{z}^{(2)}$&$L_{2x}^{(+)}, L_{3y}^{(+)}, F_{z}^{(+)}$\\
 $\Gamma_4 p_z$&+&+&-&-&-&-&+&+&$L_{1z}^{(1)}$&$L_{1z}^{(2)}$&$L_{2x}^{(-)}, L_{3y}^{(-)}, F_{z}^{(-)}$\\
 $\Gamma_5 m_y$&+&-&+&-&+&-&+&-&$F_{y}^{(1)}, L_{2x}^{(1)}$&$F_{y}^{(2)}, L_{2x}^{(2)}$&$L_{1x}^{(+)}, F_{y}^{(+)}, L_{3z}^{(+)}$\\
 $\Gamma_6 p_y$&+&-&+&-&-&+&-&+&$L_{1y}^{(1)}, L_{3x}^{(1)}$&$L_{1y}^{(2)}, L_{3x}^{(2)}$&$L_{1x}^{(-)}, F_{y}^{(-)}, L_{3z}^{(-)}$\\
 $\Gamma_7 m_x$&+&-&-&+&+&-&-&+&$F_{x}^{(1)}, L_{2y}^{(1)}$&$F_{x}^{(2)}, L_{2y}^{(2)}$&$F_{x}^{(+)}, L_{1y}^{(+)}, L_{2z}^{(+)}$\\
 $\Gamma_8 p_x$&+&-&-&+&-&+&+&-&$L_{1x}^{(1)}, L_{3y}^{(1)}$&$L_{1x}^{(2)}, L_{3y}^{(2)}$&$F_{x}^{(-)}, L_{1y}^{(-)}, L_{2z}^{(-)}$\\
\end{tabular} \end{ruledtabular} \end{table*}

The kind of exchange order (e.g., the type of magnetic order in the Heisenberg approximation) can be analyzed by considering the permutation symmetry of the magnetic moments. It accounts for a mutual orientation of magnetic moments and does not fix its global direction. Respective permutation modes [Eqn. (1)] are listed with their corresponding IRP in Table V.

\begin{table} \caption{\label{tab:table5}Permutation modes for three types of Wyckoff positions in Fe$_2$O(SeO$_3$)$_2$.}
\begin{ruledtabular} \begin{tabular}{ c|c|c|c }
  &Fe1&Fe2&Fe3\\ \hline
 $\Gamma_1$&$\vec{F}^{(1)}$&$\vec{F}^{(2)}$&$\vec{F}^{(+)}$\\
 $\Gamma_2$&$\vec{L}_1^{(1)}$&$\vec{L}_1^{(2)}$&$\vec{F}^{(-)}$\\
 $\Gamma_3 m_z$&$\vec{L}_2^{(1)}$&$\vec{L}_2^{(2)}$&$\vec{L}_1^{(+)}$\\
 $\Gamma_4 p_z$&$\vec{L}_3^{(1)}$&$\vec{L}_3^{(2)}$&$\vec{L}_1^{(-)}$\\
 $\Gamma_5 m_y$&--&--&$\vec{L}_2^{(+)}$\\
 $\Gamma_6 p_y$&--&--&$\vec{L}_2^{(-)}$\\
 $\Gamma_7 m_x$&--&--&$\vec{L}_3^{(+)}$\\
 $\Gamma_8 p_x$&--&--&$\vec{L}_3^{(-)}$\\
\end{tabular} \end{ruledtabular} \end{table}

As follows from the data given in Table V, four types of uniform collinear order can appear in Fe$_2$O(SeO$_3$)$_2$. All possible magnetic structures are shown in Fig. 6.

In the above mentioned structures the Heisenberg exchange interactions select the magnetic order parameters which are represented by their respective permutation modes [Eqn. (1)]. Anisotropic interactions lead to preferred global directions of magnetic moments (e.g., Cartesian components of the permutation modes) whereas Dzyaloshinsky-Moriya interactions (DMI), present in all types of magnetic bonds, lead to a weak non-collinearity. In a more detailed consideration we have to apply the Landau concept, in which order parameters from only one irreducible representation must be non-zero throughout a second order phase transition. The weak ferromagnetic moment may appear in the sets of order parameters belonging to either the $\Gamma_5$ or $\Gamma_7$ IRPs along the $b$ or $a$ direction, respectively (see Table IV). The largest antiferromagnetic order parameters $\vec{L}_2^{(1)}, \vec{L}_2^{(2)}, \vec{L}_1^{(+)}$ must be directed along the $a$ or $b$ axis, respectively. However, both sets of possible order parameters $\Gamma_5$ or $\Gamma_7$ are in contradiction with our susceptibility data in which a ferromagnetic response has been detected simultaneously along the $a$ direction, as well as a direction perpendicular to the $a$ axis. Such a ferromagnetic response, at $T$ close to the Neel temperature, is typical for the onset of an incommensurate order. Moreover, in our preliminary spin wave calculations the structure with both FM intralayer and interlayer interchain interactions does not show any incommensurate instability. Therefore, the antiferromagnetic order of $\Gamma_3$-type with $\vec{L}_2^{(1)}, \vec{L}_2^{(2)}, \vec{L}_1^{(+)}$ as the main order parameters has to be excluded from our consideration. Furthermore, we proceed with the $\Gamma_4$-type of antiferromagnetic order with both AFM intralayer and interlayer interchain interaction that results in an incommensurate instability along the $a$ axis and the appearance of electronic polarizations.

Our aim is to calculate the energies of the linear spin-waves in their explicit form at $\vec{k} = 0$ to elucidate the exchange interactions responsible for the activation of the zero-energy flat band modes. Additionally, we want to emphasize the results of the numerical studies of the spin-wave spectrum in Fe$_2$O(SeO$_3$)$_2$ by a comparison with analytical results. In the Heisenberg approximation for 3D long-range order we expect 16 SW branches, two of them being Goldstone modes. As follows from our preliminary model calculations the development of any incommensurate order does not sufficiently influence the spectra of high energy spin-waves. Also the model calculations show the absence of dispersion of some spin waves along the (010) and (001) directions, in spite of accounting for the 3D interchain interactions [Fig. 5(a)]. The flatness of the spin wave provides a high density of states for spin waves with energies close to $\vec{k} = 0$. Therefore, the energies of two-magnon scattering signals can be roughly estimated as a sum of magnon energies at the $\Gamma$ point. Based on the data of Table V, the magnetic Hamiltonian of Fe$_2$O(SeO$_3$)$_2$ in the nearest-neighbor Heisenberg exchange approximations at $\vec{k} = 0$ can be written in the form:

\begin{equation}
\label{eq:2}
\begin{aligned}
H=&\frac{1}{16} (j+I_l) [\vec{F}^{(+)2}+\vec{L}_1^{(+)2}-\vec{L}_2^{(-)2}-\vec{L}_3^{(-)2}]- \\
  &\frac{1}{16} (j-I_l) [\vec{L}_2^{(+)2}+\vec{L}_3^{(+)2}-\vec{F}^{(-)2}-\vec{L}_1^{(-)2}]+ \\
	&\frac{1}{4} (\frac{J}{2}+\delta) [\vec{F}^{(+)}\vec{F}^{(1)}+\vec{F}^{(-)}\vec{L}_1^{(1)}-\vec{L}_1^{(+)}\vec{L}_2^{(1)}-\vec{L}_1^{(-)}\vec{L}_3^{(1)}]+ \\
	&\frac{1}{4} (J+I_p) [\vec{F}^{(+)}\vec{F}^{(2)}+\vec{L}_1^{(-)}\vec{L}_3^{(2)}]+ \\
	&\frac{1}{4} (J-I_p) [\vec{F}^{(-)}\vec{L}_1^{(2)}+\vec{L}_1^{(+)}\vec{L}_2^{(2)}]+ \\
	&\frac{1}{2} J [\vec{F}^{(1)}\vec{F}^{(2)}+\vec{L}_1^{(1)}\vec{L}_1^{(2)}-\vec{L}_2^{(1)}\vec{L}_2^{(2)}-\vec{L}_3^{(1)}\vec{L}_3^{(2)}]
\end{aligned}
\end{equation}

Here we use the following notation: $I_l$ -- intralayer interchain exchange between Fe3-Fe3 ions, $I_p$ -- interlayer interchain exchange between Fe2-Fe3 ions. The intrachain interactions, as shown in Fig. 4(c), are given by $j$ -- vertex-vertex exchange between Fe3-Fe3 ions, $J/2+\delta$ -- spine-vertex exchange between Fe2-Fe3 ions, $J$ -- spine-vertex exchange between Fe1-Fe3 ions and the spine-spine exchange between Fe1-Fe2 ions. We take the minimal set of exchange integrals that provides 3D long-range order as well we accounts for the absence of Fe1-Fe1 and Fe2-Fe2 interactions in the nearest neighbor approximation [see Fig. 4(a)].

\begin{figure}
\label{figure6}
\centering
\includegraphics[width=8cm]{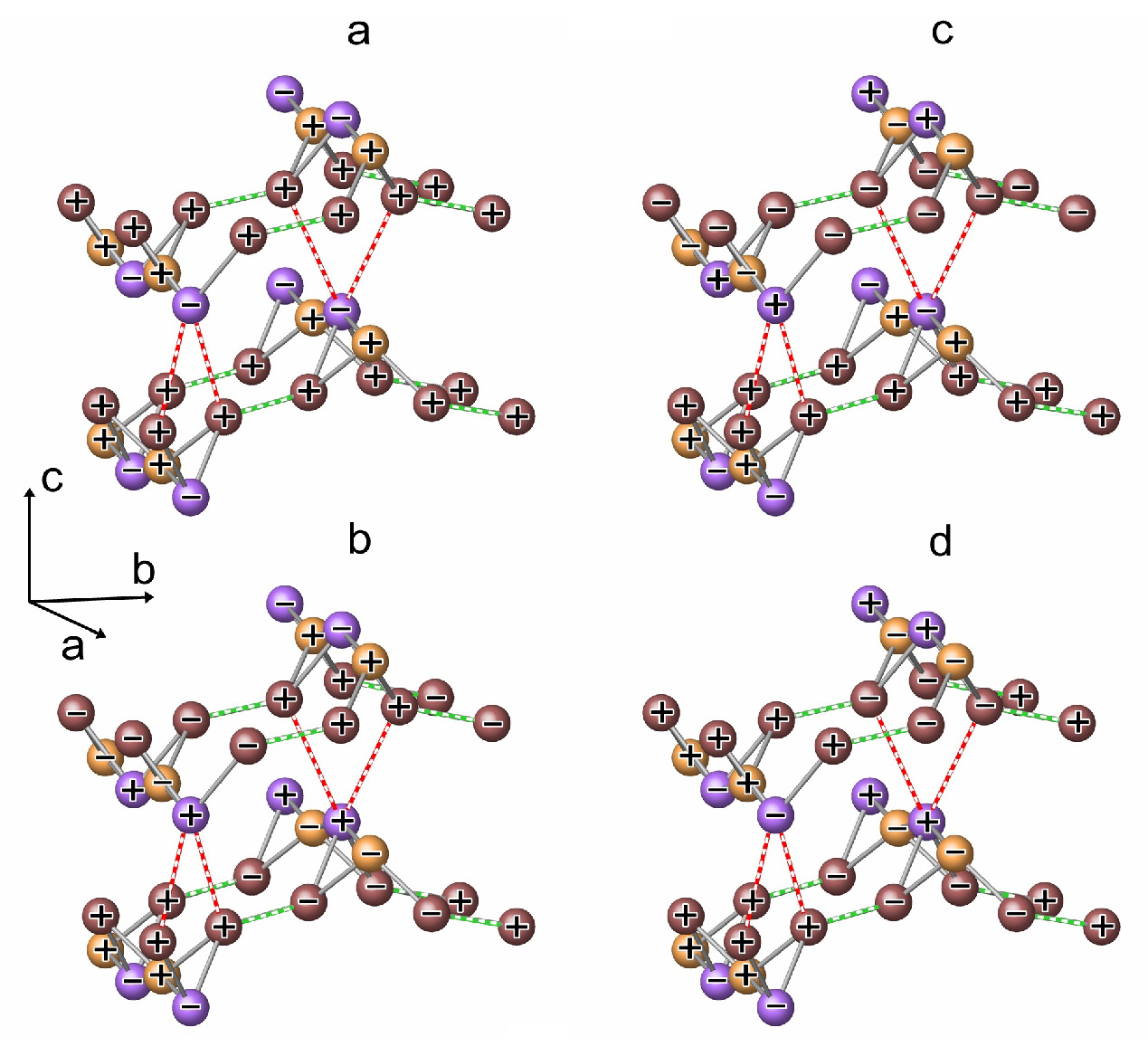}
\caption{(Color online) Sketch of the four possible uniform magnetic structures in Fe$_2$O(SeO$_3$)$_2$. The $+/-$ signs denote the mutual orientation of the Fe spins. The Fe ions in $4c$, $4d$ and $8e$ Wyckoff sites are shown in gold, violet and dark brown respectively. The intralayer interchain exchanges are marked by dashed green lines. The interlayer interactions are realized through Fe2-O-Se-O-Fe3 bonds and highlighted by dashed red lines. Fe(1)-Fe(2) interactions along the saw-tooth spine and interactions between vertices and spine are considered to be antiferromagnetic in all structures. (a) FM intralayer interchain order and AFM interlayer interchain order ($\Gamma_1$-type of ferrimagnetic order). (b) AFM intralayer interchain order and FM interlayer interchain order ($\Gamma_2$-type of antiferromagnetic order). (c) FM intralayer interchain order and FM interlayer interchain order ($\Gamma_3$-type of antiferromagnetic order). (d) AFM intralayer interchain order and AFM interlayer interchain order ($\Gamma_4$-type of antiferromagnetic order).}
\end{figure}

We proceed with equations of motion for the Cartesian components of the permutation modes [Eqn. (1)] with the following linearization. We use the $\Gamma_4$-type of antiferromagnetic uniform exchange order as a reference state to estimate energies of 14 non-Goldstone modes, or so-called exchange modes, at $\vec{k} = 0$. In the Heisenberg approximation these energies do not depend on the direction of the order parameters. For simplicity we choose the ground state with the order parameters $L_{2x}^{(1)}=4S$; $L_{2x}^{(2)}=4S$; $L_{1x}^{(+)}=-8S$, all of which belong to the $\Gamma_6$ IRP. The set of equations $i \hbar \dot{L}_{\alpha l}^{(i)} = [H,L_{\alpha l}^{(i)}]$ after linearization decomposes into four independent blocks. Every block includes the $y$- and $z$-components of the permutation modes [Eqn. (1)] from two different IRPs. The linearization procedure dictates that the IRPs of the entangled Cartesian components [Eqn. (1)] in a given block {$\Gamma_n, \Gamma_m$} are subject to a rule $\Gamma_n \otimes \Gamma_m = \Gamma_{\mathrm{GS}}$, where $\Gamma_{\mathrm{GS}}$ is the IRP of the ground state (in our case $\Gamma_{\mathrm{GS}} = \Gamma_6$). The IRP content of the block defines the symmetry of respective modes. The blocks with IRP {$\Gamma_2, \Gamma_5$} and {$\Gamma_3, \Gamma_8$} include the $z$- and $y$-components of the main order parameters, respectively; therefore, the solution of the respective dynamic matrices must include the energies of the Goldstone modes which are zero at $\vec{k} = 0$. The sets of equations in the {$\Gamma_1, \Gamma_6$} and {$\Gamma_4, \Gamma_7$} blocks describe only energies of exchange modes. The {$\Gamma_1, \Gamma_6$} set of equations has the form:

\begin{widetext}
\begin{equation}
\hat{M}_{16}L_{16} =
\begin{pmatrix}
i\omega & -A & 0 & B & 0 & C & 0 & 0 \\
A & i\omega & B & 0 & C & 0 & 0 & 0 \\
0 & B & i\omega & -D & 0 & -F & 0 & 0 \\
B & 0 & D & i\omega & F & 0 & 0 & 0 \\
0 & -2C & 0 & 2F & i\omega & X & 0 & 0 \\
-2C & 0 & -2F & 0 & Y & i\omega & 0 & 0 \\
0 & 0 & 0 & 0 & 0 & 0 & i\omega & P \\
0 & 0 & 0 & 0 & 0 & 0 & Q & i\omega \\
\end{pmatrix}
\cdot
\begin{pmatrix}
L_{1y}^{(1)} \\
L_{2z}^{(1)} \\
L_{1y}^{(2)} \\
L_{2z}^{(2)} \\
F_{y}^{(-)} \\
L_{1z}^{(-)} \\
L_{2y}^{(+)} \\
L_{3z}^{(-)} \\
\end{pmatrix}
=
\begin{pmatrix}
0 \\
0 \\
0 \\
0 \\
0 \\
0 \\
0 \\
0 \\
\end{pmatrix}
\end{equation}
\end{widetext}

Here: $A = (J-2\delta)S$, $B = 2JS$, $C = (J/2+\delta)S$, $D=(4J+2I_p)S$, $F = (J-I_p)S$, $X = (J/2-\delta +I_p + 2I_l)S$, $Y = -(J/2-\delta + I_p)S$, $P = (J/2 - \delta + I_p -2j)S$, $Q = -(J/2-\delta + I_p - 2j+2I_l)S$. The specific topology of nearest-neighbor exchange bonding as well as the minimal set of exchange integrals leads to an additional degeneracy of spin waves. It is manifested by $\Hat{M}_{47} = \Hat{M}_{16}$ in the {$\Gamma_4, \Gamma_7$} set of equations $\Hat{M}_{47}L_{47} = 0$. The row $L_{47}$ includes next Cartesian components of [Eqn. (1)]: $L_{2y}^{(1)}, L_{1z}^{(1)}, L_{2y}^{(2)}, L_{1z}^{(2)}, L_{1y}^{(+)}, F_{z}^{(-)}, L_{2z}^{(+)}, L_{3y}^{(-)}$. Similarly, we have $\hat{M}_{25} = \hat{M}_{38}$ for {$\Gamma_2, \Gamma_5$} and {$\Gamma_3, \Gamma_8$} sets of equations $\hat{M}_{25}L_{25} = 0$ and $\hat{M}_{38}L_{38} = 0$. The rows $L_{25}$ and $L_{38}$ include the next Cartesian components of [Eqn. (1)]: $F_{y}^{(1)}, L_{3z}^{(1)}, F_{y}^{(2)}, L_{3z}^{(2)}, F_{y}^{(+)}, L_{1z}^{(-)}, L_{3z}^{(+)}, L_{2y}^{(-)}$ and $L_{3y}^{(1)}, F_{z}^{(1)}, L_{3y}^{(2)}, F_{z}^{(2)}, L_{1y}^{(-)}, F_{z}^{(+)}, L_{2z}^{(-)}, L_{3y}^{(+)}$, respectively. The matrix $\hat{M}_{25}$ has exactly the same form as the matrix $\hat{M}_{16}$ in which we should replace $\hat{M}_{25} = \hat{M}_{16} (X \rightleftarrows -Y, F \rightarrow (J+I_p)S)$.

One of the distinct features of the spectrum are the modes arising separately in every $8 \times 8$ block and having the same energy at the $\Gamma$ point; $S(-PQ)^{1/2}$. As follows from numerical calculations these so-called ``separate'' modes evidence some dispersion and splitting into double degenerate SW along the (100) direction but they exhibit fourfold degeneracy and dispersionless behavior along the (001) direction and very small dispersion along the (010) direction. Respective excitations are located on the vertex ions (e.g., on the Fe3 sublattice) and can propagate only along the saw-tooth chain in spite of presence of the 3D interchain interactions $I_l$ and $I_p$. In the case of the classical $\Delta$ chain with $\delta = J/2$, and $j = I_p = 0$ these modes are zero energy dispersionless modes~\cite{nakamura-96, sen-96}.

The solutions of the remaining part of every $6 \times 6$ block do not contain the intrachain vertex-vertex interaction $j$. Considering $\delta = I_p = I_l = 0$, one can estimate a few separate scales of the SW spectrum of the isolated saw-tooth chain which are given by the energies $2.5SJ$ and $S \vert J/2-2j \vert$ of the two highest modes at the $\Gamma$ point and the two zero-energy modes, consisting of the Goldstone mode and the flat-band mode. The presence of four saw-tooth units in the unit cell of Fe$_2$O(SeO$_3$)$_2$ makes all modes fourfold degenerate.

In the case of a 2D $ab$-layer of saw-tooth chains with $\delta = 0$ interacting through the $I_l$ exchange the double degenerate zero-energy flat-band modes survive along the (010) directions. In that case, there is an insufficient renormalization of the highest modes energies $0.5S (25J^2+ 4 JI_l)^{1/2}$ and $S\sqrt{(J/2-2j)(J/2-2j+2I_l)}$. The intralayer exchange $I_l \neq 0$ by itself does not create an additional gap and the SW spectrum of every 2D layer of interacting saw-tooth chains includes two zero energy flat-band modes. Four degenerate Goldstone-like modes exist along the (100) direction in the every $ab$-layer.

In the case of $\delta \neq 0$ for an isolated saw-tooth unit ($I_p = I_l = 0$) one out of two zero-energy flat-band modes obtains the gap $16S \delta J (5J + 6\delta)^{-1}$. The solutions for SW energies at the $\Gamma$ point are the same for every one out of four blocks of equations (e.g., every SW branch is fourfold degenerate). For the case $\delta \ll J$ and $I_l \ll J$ these solutions have the form:

\begin{equation}
\begin{aligned}
\omega_I &= 0; \\
\omega_{II} &\approx S \sqrt{\frac{32 \delta J}{(5J+6\delta)^2}[8 \delta J - I_l (J-2 \delta)]}; \\
\omega_{III} &= S \sqrt{(J/2- \delta - 2j)(J/2- \delta -2j + 2I_l)}; \\
\omega_{IV} &\approx \frac{S(5J+6\delta)}{2}\left( 1-\frac{32\delta J}{(5J+6\delta)^2}+\frac{2I_l(J-2\delta)}{(5J+6\delta)^2} \right)
\end{aligned}
\end{equation}

These solutions highlight minor contributions of the intralayer interchain interaction $I_l$ for the gap formation. The interlayer interchain interaction $I_p \neq 0$ together with $\delta \neq 0$ both activate zero-energy flat-band modes at the $\Gamma$-point in the {$\Gamma_1, \Gamma_6$} and {$\Gamma_4, \Gamma_7$} sets of equations. However, the exact analytical solutions of the {$\Gamma_2, \Gamma_5$} and {$\Gamma_3, \Gamma_8$} sets of equations each hold the zero frequency mode, providing at $\vec{k} = 0$ the onset of two Goldstone modes expected for the 3D collinear antiferromagnet.

In the case of interlayer exchange $I_p$ the four-fold degeneracy decreases into a two-fold one. Surprisingly, the ``separate'' modes remain unsplit. The splitting of the highest modes depends linearly on the magnitude of $I_p$. The activated zero-energy modes experience a similar splitting. It can be illustrated under the approximations $\delta = I_l = 0$ by the following solutions:

\begin{equation}
\begin{aligned}
\omega_+^{(16,47)}(0) &\approx S \frac{(5J+2I_p)(1+x/2)}{2}; \\
\omega_{I-}^{(16,47)}(0) &\approx \frac{20SJI_p}{(5J+2I_p)}; \\
\omega_{II-}^{(16,47)}(0) &= 0; \\
\omega_+^{(16,47,25,38)}(0) &\approx S \frac{(5J+2I_p)(1+x/10)}{2}; \\
\omega_-^{(25,38)}(0) &\approx \frac{4SJI_p}{(5J+2I_p)}; \\
\omega_{Gold}^{(25,38)}(0) &= 0; \\
\omega_{sep}^{(16,47,25,38)}(0) &= S (J/2-2j+I_p); \\
x &= \frac{80JI_p}{(5J+2I_p)^2} \ll 1;
\end{aligned}
\end{equation}

Here, the upper indices enumerate solutions of the respective set of equations. Numerical calculations establish a decisive contribution of interlayer interchain interaction $I_p$ for the mode splitting and the formation of gaps, whereas the non-zero parameter $\delta$ by itself does not lift the four-fold degeneracy but induces a gap for $\omega_{II-}^{(16,47)}(0)$ modes. The solutions [Eqn. (5)] can be used for rough estimations of the observed energy scales of the two-magnon Raman peaks in Fe$_2$O(SeO$_3$)$_2$.

We proceed with the approximation that the two-magnon Raman peaks should be localized at the SW energies close to $\vec{k} = 0$ where the flatness of the SW spectrum provides the highest density of states. The exchange mechanism of two-magnon scattering supposes that the Raman cross-section in parallel polarization must contain magnon branches with the same symmetry. In our case the two-magnon scattering is allowed on two different SW branches which must originate from the same symmetry blocks. In particular, the peaks at 600 cm$^{-1}$ and 523 cm$^{-1}$ can be assigned to such a composite of high energy and activated low energy magnons $\omega_{+}^{(16,47)}(0) + \omega_{II-}^{(16,47)}(0) \approx 600$ cm$^{-1}$ and $\omega_{+}^{(25,38)}(0) + \omega_{-}^{(25,38)}(0) \approx 523$ cm$^{-1}$. The low lying two-magnon signal might originate only from activated flat-band zero energy magnons $2\omega_{-}^{(16,47)}(0) \approx 54$ cm$^{-1}$. The specific saw-tooth geometry, unlike a linear chain AFM, enables the scattering process in crossed polarizations. In the Loudon-Fleury formalism the Hamiltonian for two-magnon scattering has the form~\cite{fleury-68, cepas-08}:

\begin{equation}
H_R = \sum_{<ij\alpha \beta>} (\vec{e}_{\mathrm{in}} \vec{r}_{i\alpha j\beta}) (\vec{e}_{\mathrm{out}} \vec{r}_{i\alpha j\beta}) (\vec{s}_{i\alpha} \vec{s}_{j\beta}),
\end{equation}
where: $\vec{e}_{\mathrm{in}}$, $\vec{e}_{\mathrm{out}}$ are polarizations of the incident and scattered light; $i,j = {1,2,3}$ denote Fe sublattices; $\alpha, \beta$ are the numbers of the Fe ions in the respective sublattices; $\vec{r}_{i\alpha j\beta}$ is the bond vector connecting nearest-neighbor $i\alpha$ and $j\beta$ sites. In Fe$_2$O(SeO$_3$)$_2$ the spine-vertex bonds Fe1-Fe3 and Fe2-Fe3 have $z$-components with a length roughly equal to 0.86 of the Fe1-Fe2 bond along the chain direction. Therefore, from a lattice topology point of view, the $xx$- and $xz$ Raman signals may have comparable intensity. The two-magnon scattering in $xz$-cross polarization probes a composite of the {$\Gamma_3,\Gamma_8$} and {$\Gamma_4,\Gamma_7$} types of magnons. The peak at 163 cm$^{-1}$ can be assigned to the scattering on the combination of the ``separate'' magnons and one of the activated low energy magnons $\omega_{sep}^{(38)}(0) + \omega_{I-}^{(47)}(0) = 163$ cm$^{-1}$.

Using the assignments given above of four two-magnon modes and the analytical solution of the equations $\hat{M}_{38}L_{38} = 0$ and $\hat{M}_{47}L_{47} = 0$ for magnon energies one can roughly estimate the five exchange parameters $J$, $I_p$, $j$, $I_l$, and $\delta$ in Fe$_2$O(SeO$_3$)$_2$. We additionally use the inequality $j > I_p$ as a condition that follows from the lengths of Fe3-O-Se-O-Fe3 (4.32 \AA) and Fe3-O-Se-O-Fe2 (4.64 \AA) pathways and the condition that the values of $j$ and $I_p$ must have the same sign. The last two conditions remove the uncertainty in the types of possible magnon contribution into two-magnon signals. The restored constants in cm$^{-1}$ are: $J = 75.84$; $I_p = 9.18$; $j = 12.1$; $I_l = 10$, and $\delta = 3.77$. Here we take the value $I_l = 10$ cm$^{-1}$ as the lowest value at which the inequality $j > I_p$ is fulfilled. The magnon energies at $\Gamma$ point in cm$^{-1}$ are:

\begin{equation}
\begin{aligned}
\omega_+^{(16,47)}(0) &= 573 \\
\omega_{I-}^{(16,47)}(0) &= 94.6 \\
\omega_{II-}^{(16,47)}(0) &= 27 \\
\omega_+^{(25,38)}(0) &= 518 \\
\omega_{-}^{(25,38)}(0) &= 5 \\
\omega_{Goldst}^{(25,38)}(0) &= 0 \\
\omega_{sep}^{(16,47,25,38)}(0) &= 68.4
\end{aligned}
\end{equation}

The estimated constants serve as an ``order-of-magnitude approximation'' due to the very rough estimation of the two-magnon density of states, as well as due to the neglect of the two-magnon signal linewidth.

\end{document}